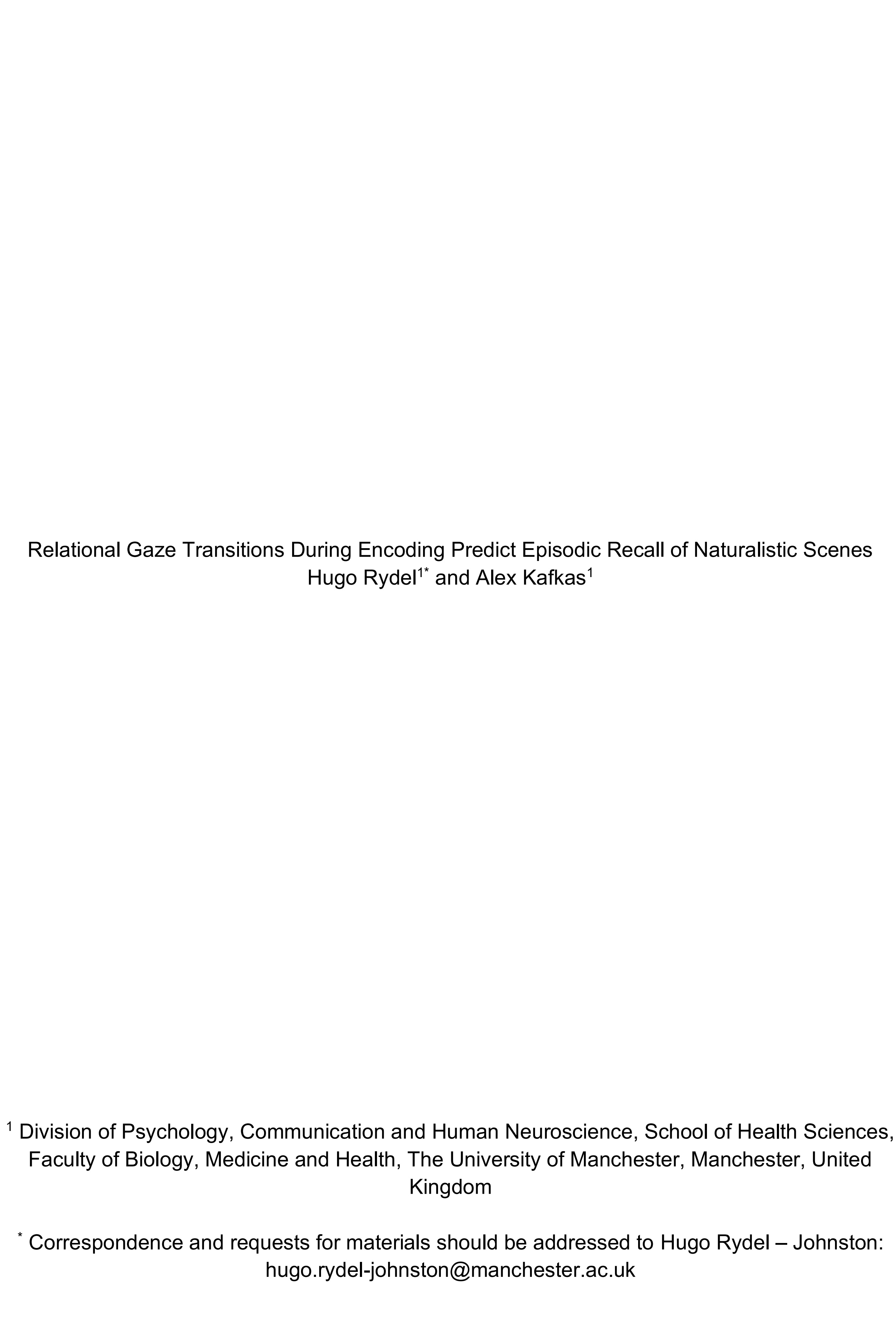

# Relational Gaze Transitions During Encoding Predict Episodic Recall of Naturalistic Scenes

Hugo Rydel[1*] and Alex Kafkas[1]

[1] Division of Psychology, Communication and Human Neuroscience, School of Health Sciences, Faculty of Biology, Medicine and Health, The University of Manchester, Manchester, United Kingdom

[*] Correspondence and requests for materials should be addressed to Hugo Rydel – Johnston: hugo.rydel-johnston@manchester.ac.uk

## Abstract

Remembering a visual scene requires organizing distinct details into a cohesive event. This study investigates whether relation-guided gaze transitions provide a behavioural marker of this cognitive organization during episodic encoding and retrieval. By applying scene graph annotations to eye-tracking data, we measured whether gaze moved between objects that were meaningfully related within complex scenes. This approach allowed us to quantify relational scanning within naturalistic environments, moving beyond prior methods that relied on simplified displays or isolated relation types. Participants showed above-chance relational gaze during both initial viewing and blank-screen retrieval, indicating that gaze actively tracks scene structure during first viewing and at recall. Additionally, relational scanning at encoding predicted subsequent free recall of both object and relational details, even after accounting for salience, fixation frequency, meaning, and image-level differences. In contrast, relational scanning at retrieval did not predict recall success, suggesting that relational gaze is most functional to memory during its formation. Together, these findings show that relational gaze can be measured in complex scenes and may serve as a marker of episodic encoding during natural visual exploration.

## Introduction

Episodic memory involves the conscious recollection of previous experiences together with their time, place, and context (Tulving, 1972, 2002). Because visual experience unfolds through eye movements, gaze provides one route through which information may enter episodic memory. Fixations provide brief temporal windows in which detailed visual information can be extracted, while saccades rapidly reposition gaze toward new regions of a scene, creating a sequence of sampled scene events (Rayner, 1998). Accordingly, gaze behavior at both encoding and retrieval reliably associates with episodic memory performance (Fehlmann et al., 2020; Lucas et al., 2023), offering insight into how memory may be constructed and retrieved in real time (Ryan & Shen, 2020; Lucas et al., 2023). However, while this literature establishes a clear association between gaze and memory outcomes, the specific processes that allow a sequence of eye movements to build a coherent memory representation remain less clear. Determining how gaze facilitates episodic remembering is therefore an important step in moving beyond descriptive gaze-memory accounts.

Attentional selection offers one intuitive account of the gaze-memory link. Because visual processing capacity is limited (Kastner & Ungerleider, 2000), selective attention helps determine which visual inputs receive the resources needed to support long-term encoding (Chun & Turk-Browne, 2007). Gaze may therefore support memory by selecting locations or objects, allowing them to be attended, and thus increasing their likelihood of encoding. Existing accounts differ in what explains a region's attentional priority. Bottom-up salience accounts emphasise low-level visual contrast (Itti & Koch, 2000), whereas meaning-map accounts emphasise the semantic informativeness of scene regions (Henderson et al., 2019). Despite their differences, both salience and meaning-map approaches can predict where viewers look in scenes (Parkhurst et al., 2002; Henderson et al., 2019), with meaning-based accounts offering uniquely strong predictions of gaze attention (Henderson & Hayes, 2017, 2018). By guiding fixation allocation, these models suggest that regions high in visual salience or semantic meaning should be more reliably remembered due to preferential attention-driven processing.

However, memorability research suggests that attentional selection and episodic memory are partly dissociable (Rust & Mehrpour, 2020). For example, in a drawing-based recall task using real-world scenes, Bainbridge et al. (2019) found that salience was only marginally predictive of delayed recall, while meaning predicted no significant recall differences. Direct tests of attentional capture support the same distinction. Sasin and Fougnie (2021) found that while salient distractors captured more attention during visual search than equivalent non-salient distractors, this attentional advantage did not translate into better memory. Consequently, evidence from both visual search and scene memorability suggests that simply attending to salient or meaningful elements is not enough to guarantee their retention. This dissociation may arise because attention selection accounts focus primarily on which individual scene elements receive processing, whereas episodic memory also depends on how those elements are organised and bound within a broader episode, including what happened, where, and when (Tulving, 1984). Thus, what attentional accounts may be missing is the structural framework needed to explain how individually attended elements are integrated into a unified representation. To understand how momentary visual details become a cohesive episode, it may therefore be useful to examine the binding processes that support this organization within memory.

Relational accounts offer a possible explanation for how this organization might occur. During viewing, observers form a coherent model of a scene by identifying how objects, agents, and actions interact (Radvansky & Zacks, 2014; Zacks et al., 2007). This “event model” is built upon spatial, functional, and semantic relations, which specify how separate visual details belong together in the scene (Zwaan & Radvansky, 1998; Võ et al., 2019). By establishing these structural links, relations allow newly attended items to be incorporated into the ongoing scene context (Loschky et al., 2020), thus offering a way through which momentary perceptual details can be transformed into a unified representational structure. In this context, relations may be relevant to the gaze-memory link because they describe how momentary visual samples can be situated within the broader episode being encoded. A relational perspective therefore offers a useful way to consider how such episodes might later be stored and contextualised in memory.

This proposed framework aligns closely with relational memory theory, which proposes that episodic binding within the hippocampus depends on linking elements through their relations (Konkel & Cohen, 2009; Eichenbaum, 2017). Further evidence shows that relational binding is enhanced when relations fit their broader scene context (Davachi & Wagner, 2002; Staresina et al., 2009), supporting the role relations may play in contextual integration. Taken together, relations may provide the organising structure through which attended elements are incorporated into a cohesive memory representation. However, while relational memory theory explains the kind of representations that support episodic recall, it does not explain the moment-by-moment viewing behaviours that might support this binding process during scene exploration. Without this link, this model remain abstracted from how information is acquired during viewing. Thus, identifying real-time behavioural markers of relational integration is needed to show how cognitive models of memory may relate to natural viewing behaviour.

Gaze transitions offer one way this relational encoding process could occur. Because observers sample only part of a scene at any moment, relational encoding often depends on how information is accumulated across successive eye movements (Henderson & Hollingworth, 1999). As viewing proceeds, the active event context can guide gaze toward likely relational partners (Radvansky & Zacks, 2014; Võ & Wolfe, 2013), for instance, moving from a hand to the hammer it holds, to the nail it is striking. Through these gaze transitions, information from an initially fixated element can remain active in visual working memory while a related element is sampled next (Hollingworth & Henderson, 2002; Hollingworth, 2004). In this way, gaze transitions can bridge the temporal gap between related scene elements, enabling them to be processed (Henderson & Hollingworth, 2003) and potentially bound together. Because successive fixations maintain this working memory overlap, they theoretically provide the temporal conditions under which relational binding might occur. In effect, such relational transitions may transform an isolated act of attentional selection into a period of contextual integration, suggesting that the specific relations transitioned by gaze could influence how effectively an episodic scene is constructed. Thus, testing whether viewers systematically make relation-guided transitions can help determine whether this behaviour supports episodic encoding.

Empirical evidence from scene viewing supports this proposed mechanism by showing that gaze tends to follow relational structure. In object-based scenes, viewers make relation-sensitive transitions between objects that are semantically connected or that occupy predictable spatial locations (Hwang et al., 2011; Wu et al., 2014). This pattern also extends to relations that organise ongoing events, such as functional relations between tools and the objects they act upon (Castelhano & Witherspoon, 2016; Federico & Brandimonte, 2019), and social relations between interacting individuals (Castelhano et al., 2007; Skripkauskaite et al., 2023). These findings show that when specific relations are isolated experimentally, visual sampling can align closely with the relational structure of a scene. Yet, because real-world environments contain multiple, overlapping objects and relations that compete for attention, isolated displays may not capture how gaze operates in everyday settings. Therefore, testing whether relation-sensitive scanning persists amid the visual competition of naturalistic scenes, and whether it predicts downstream memory benefits, would help confirm its generalisability.

Gaze at retrieval provides a related but more cautious extension of these predictions. If relational transitions during viewing help form contextual bindings, relational transitions at retrieval may help reinstate this context once the scene is no longer visible. This possibility is supported by the “looking at nothing” phenomenon, where individuals spontaneously direct their gaze toward empty spatial locations associated with a previously viewed scene (Ferreira et al., 2008). Building on this spatial phenomenon, behavioural evidence indicates that gaze may help recover relational information. For instance, when a cue reinstates a learned association, viewers rapidly bias their gaze toward the item that completes that relation among distractors (Hannula et al., 2007; Chua et al., 2012). Furthermore, more recent EEG work shows that relationally guided gaze transitions are accompanied by neural memory reactivation signatures (Nikolaev et al., 2025), suggesting that relation-guided looking may support the reinstatement of stored event representations. Nevertheless, this evidence primarily links retrieval gaze to memory through cue-driven associative recognition or probed visual search, tasks which inherently require viewers to verify or locate a specific target. It remains unclear whether the same functional benefit extends to tasks with different demands, such as the richer, self-generated recall of naturalistic scenes. Investigating this unconstrained, generative recall would clarify whether retrieval gaze actively contributes to episodic reconstruction, or if its utility depends partly on how memory is probed.

Testing these predictions requires a way to quantify relational gaze transitions in naturalistic scenes. This has been difficult because previous work has often made relations measurable by isolating relation types or simplifying the visual display. Such approaches are useful for showing that gaze can follow semantic, spatial, functional, or social relations, but they leave open whether gaze is biased toward relations when multiple object pairs and relation types coexist within the same scene. A stronger test therefore requires image-specific relational maps that preserve naturalistic scene structure while specifying which objects are meaningfully related. Recent scene graph datasets provide this possibility. In particular, the Synthetic Visual Genome (SVG) provides naturalistic scene images with segmented objects and labelled semantic relations between object pairs (Park et al., 2025). These annotations allow fixations to be mapped onto object locations and successive object-to-object transitions to be classified according to whether they follow an annotated relation. By using these datasets, relation-sensitive gaze can be quantified without sacrificing naturalistic visual complexity. This approach thus offers a practical way to translate theoretical claims about relational integration into testable predictions in more ecologically valid settings.

The present study used this SVG-based approach to test whether relational gaze transitions occur systematically during naturalistic scene encoding and retrieval, and whether they are associated with episodic recall. Participants first viewed SVG-derived scenes during encoding and later retrieved each scene during a looking-at-nothing phase followed by written free recall. Eye movements were mapped onto SVG object and relation annotations, allowing successive gaze transitions to be classified according to whether they followed the relational structure of each scene. For each image, we calculated a chance-corrected relational score indicating whether object-to-object transitions followed annotated relations more often than expected from the objects fixated on that trial. Recall responses were scored for correctly remembered object and relational details, allowing us to test whether relational gaze predicted both object memory and relational memory. We predicted that gaze would show above-chance relational scores during both encoding and retrieval. We further predicted that higher relational scores would be associated with better subsequent recall, even after accounting for salience, fixation frequency, and between-image differences. In this way, the study tested whether relational gaze generalises beyond isolated relation types to the richer relational structure of naturalistic scenes, and whether this relational bias predicts episodic recall beyond simpler attentional accounts.

## Methods

### Participants

Twenty-eight psychology students from the University of Manchester were recruited for the study. All participants were native English speakers with normal or corrected-to-normal vision, and reported no history of neurological or psychiatric disorders. The mean age of the sample was 19.76 years (*SD* = 1.30 years), and 5 participants were male. Following data collection, one participant requested their data be withdrawn, and two were excluded due to eye-tracking calibration failures, leaving a final sample of 25 participants for all subsequent analyses.

### Ethics approval and informed consent

Ethical approval was granted by the University of Manchester Research Ethics Committee (Ref: 2025-24187-44486). All procedures were conducted in accordance with the Declaration of Helsinki and relevant institutional guidelines. All participants provided written informed consent before taking part.

### Design

The experiment employed a within-participant design consisting of three phases: encoding, distractor, and retrieval. The primary dependent variables were the proportions of scene elements correctly recalled, categorized into total recall, relational recall, and object recall. The primary independent predictors were relational scores at encoding and retrieval, indexing the degree to which a participant's gaze sequence transitioned between relationally linked objects beyond chance. Trial-level covariates included fixation count and salience at relational fixations.

### Stimuli and Materials

Stimuli were sampled from the SVG-SG split of the Synthetic Visual Genome (SVG) dataset (Park et al., 2025). To ensure suitable visual and relational complexity, the initial image pool was filtered to include only images containing 10 to 30 objects, at least 10 annotated relations, and a minimum segmentation mask coverage of 75%, computed from the SVG segmentation annotations. To increase the likelihood of successful recall, image memorability was estimated using ResMem, a deep learning model trained to predict human recognition of images after a short delay (https://brainbridgelab.uchicago.edu/resmem/; Needell & Bainbridge, 2022). Applying a mean ResMem memorability threshold of ≥ 0.75 (out of 1) retained the top 50% most memorable images from the dataset.

Next, an automated pre-filtering phase used OpenAI's GPT-5.4 model (OpenAI, 2026) to score the remaining images across four criteria: Core Interaction Counts (CIC), Object Separability (SEP), Scene Dynamics (DYN), and Image Quality (QLT). Images advanced only if they scored ≥ 2 (out of 3) on CIC, ≥ 1 (out of 2) on both SEP and DYN, and 1 (out of 1) on QLT (see Appendix B for the specific prompts, scoring rubrics, and criteria definitions). A final manual screening eliminated images that were too semantically similar or contained confounding artifacts like watermarks or visible text. This yielded a final set of 30 stimulus images, which were subsequently resized and letterboxed to a uniform resolution of 1024 x 768 pixels. For each image, two distinct four-option multiple-choice engagement questions were manually written by the experimenter. These questions were designed to probe different aspects of scene content, such as actions, object identities, counts, or spatial relations, and each included one correct answer and three distractors (see Appendix G, Table 1 for examples of encoding questions).

SVG-derived object segmentation masks defined the structural Areas of Interest (AOIs; see Figure 1B), while SVG-derived relations were used to construct a binary AOI × AOI relation matrix indicating the presence (1) or absence (0) of a semantic link between objects (see Figure 1C). Additionally, to account for low-level visual guidance, a spectral-residual salience map (Hou & Zhang, 2007) was computed for each image using the OpenCV Python library (https://pypi.org/project/opencv-contrib-python). The resulting maps were smoothed with a Gaussian kernel, $\sigma = 3$, and normalized to produce peaked probability distributions centered on visually salient objects.

**Figure 1**

*Representative Stimulus with SVG-Derived Object Masks and Object Relations Mapped*

**A. Example scene stimulus**

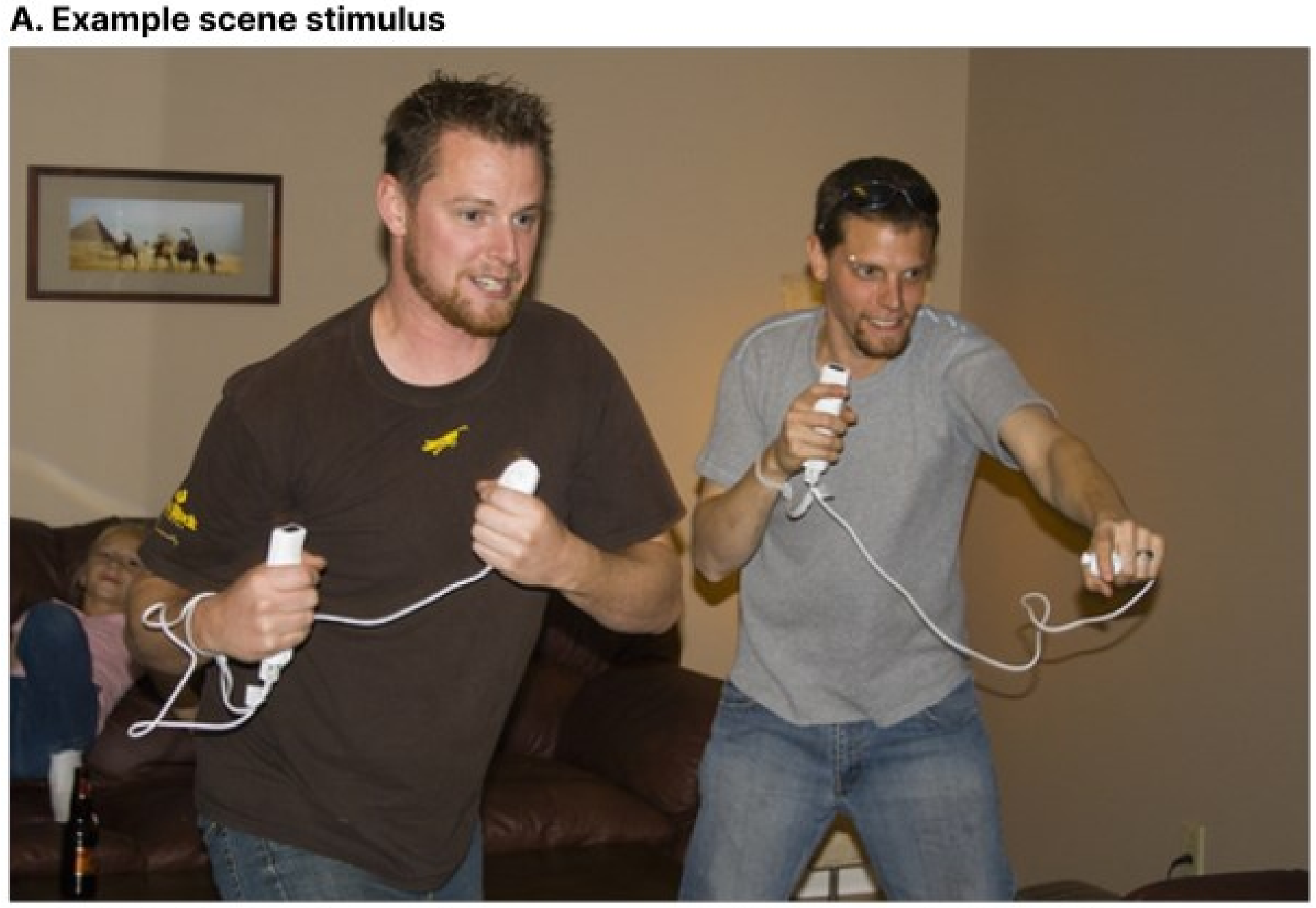

**B. SVG-derived object masks**

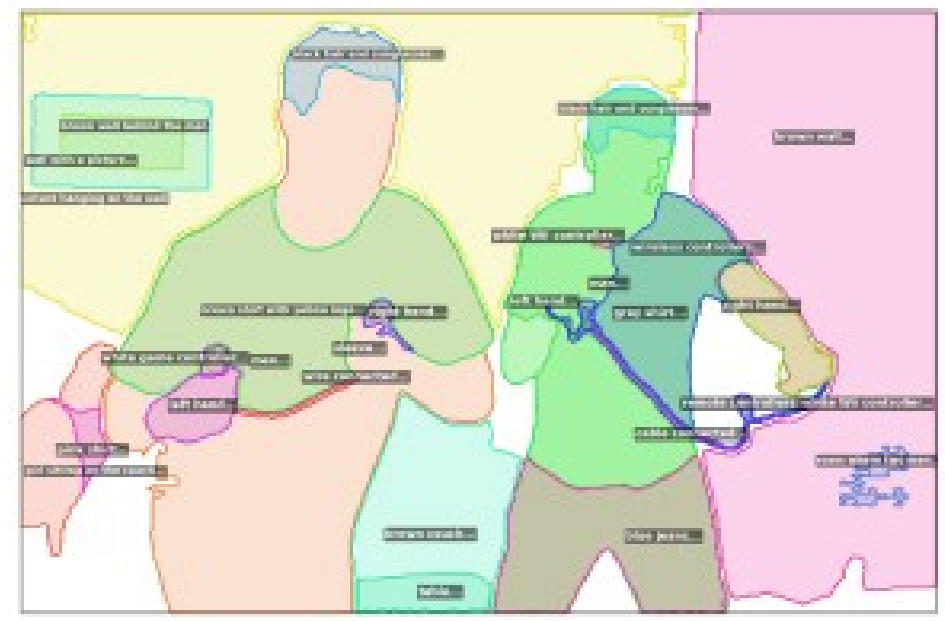

**C. SVG-derived object relations**

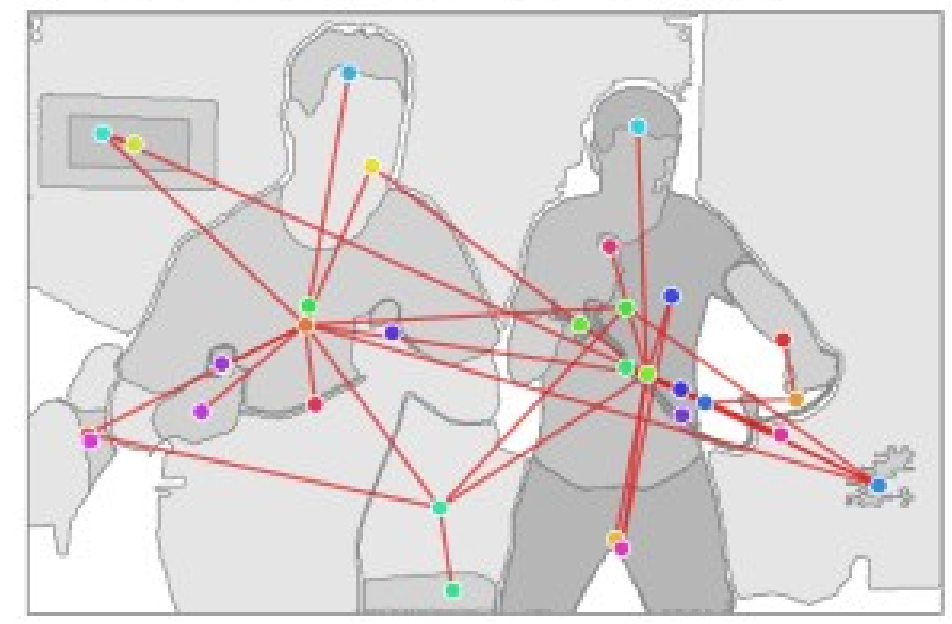

*Note.* Panel A shows a representative stimulus image used in the experiment; Panel B shows SVG-derived object masks used as AOIs; Panel C visualizes the binary AOI × AOI relation matrix, with red lines indicating existing relations between objects and color-coded circles indicating the object centers.

## Apparatus

The experiment was programmed and presented using E-Prime 3.0 (Psychology Software Tools). Stimuli were displayed on a 24-inch Dell monitor at a resolution of 1920 × 1080 pixels. Participants were seated approximately 60 cm from the screen with their head stabilized by a chinrest. Eye movements were recorded monocularly from the left eye at a sampling rate of 1000 Hz using an EyeLink 1000 Plus eye tracker (SR Research).

## Procedure

After providing informed consent, participants were calibrated using a standard 9-point eye-tracker procedure. The experiment then proceeded across three phases: encoding, distractor, and retrieval (see Figure 2 for an overview of the trial sequences).

During the encoding phase, participants were explicitly instructed to memorize 30 scene images and their associated cue words. To ensure sustained engagement and provide adequate exposure, each of the 30 scenes was presented twice in a randomized order, resulting in 60 total encoding trials. Each trial began with a 500-ms fixation cross, followed by the scene's unique cue word for 2 seconds, and the scene image itself for 5 seconds, during which eye movements were recorded. Immediately after the image disappeared, participants answered a self-paced, four-option multiple-choice engagement question (4AFC) about the scene by selecting one of the four response options via keypress. A different question was asked during the first versus the second viewing of each image (e.g., viewing one: *"How many men are playing in the scene?"*; viewing two: *"In which hand(s) are the men holding the controllers?"*). Following the encoding phase, participants completed a 5-minute arithmetic distractor task to suppress working memory rehearsal.

During the retrieval phase, participants were instructed to vividly recall the encoded scenes. Each retrieval trial commenced with a 500-ms fixation cross and the presentation of a previously studied cue word for 2 seconds. This was followed by a 7-second blank screen, during which participants were asked to mentally reinstate the cued image while their eye movements were recorded. Finally, a text box appeared on screen, and participants were given unlimited time to type out as many details as they could remember from the scene.

**Figure 2**

*Overview of the Encoding and Retrieval Trial Sequences*

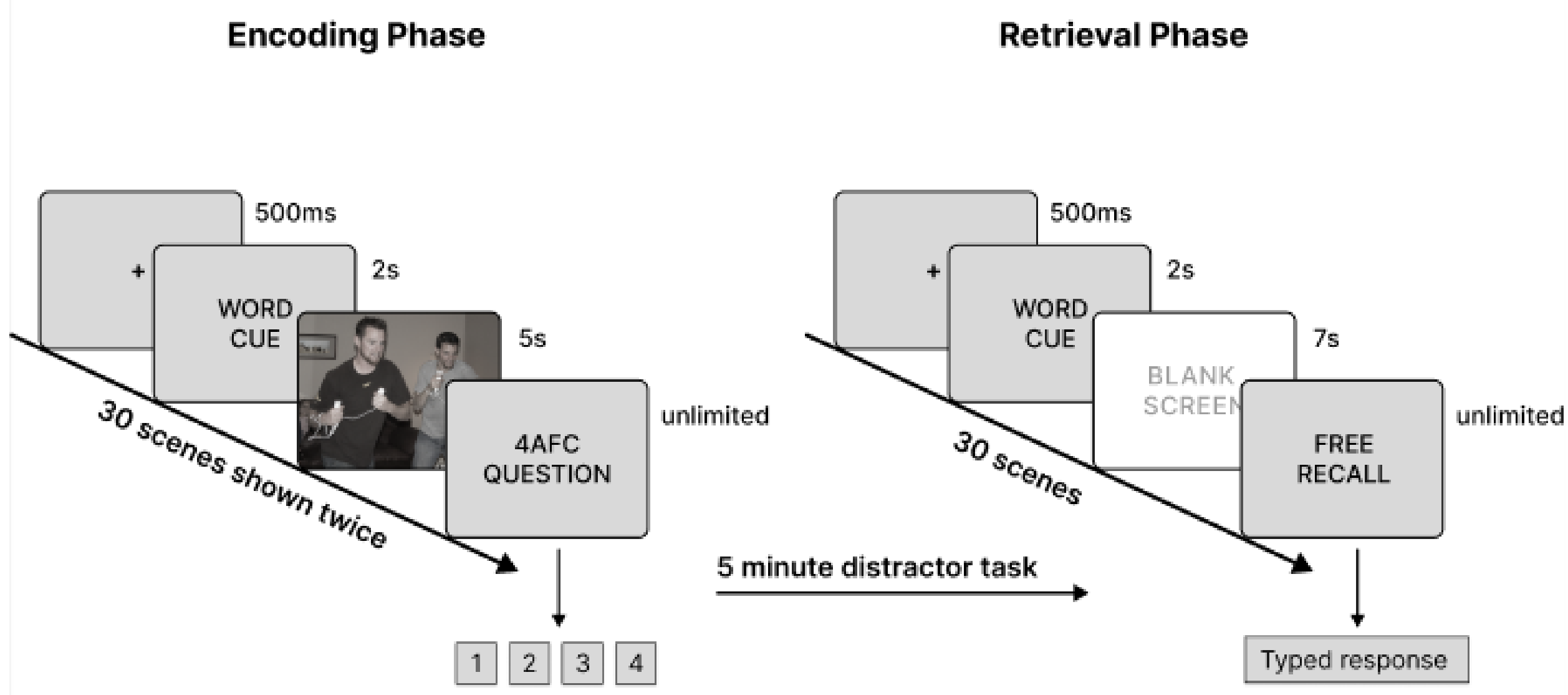

*Note.* 4AFC = four-alternative forced-choice. Timing labels are shown in milliseconds (ms) or seconds (s).

**Data Analysis**

Raw eye-tracking data were converted into trial-level relational scanning scores by assigning fixations to SVG-defined object areas of interest (AOIs). A fixation was first assigned to an object if it fell within that object's segmentation mask. If it fell outside all masks, it was assigned to the nearest object centroid only if it lay within 50 px of that centroid at encoding, or 100 px at retrieval; otherwise, it remained unassigned. Within each trial, fixations were ordered by onset time, unassigned fixations were excluded, and consecutive fixations on the same object were collapsed so that the resulting sequence reflected distinct successive object visits (e.g., a raw fixation sequence on objects A → A → B → A → C → C would be reduced to A → B → A → C).

Relational scanning was quantified from these object sequences using the SVG scene graph. For each trial, an adjacent transition was scored as relational when the two successive objects were linked in the predefined binary AOI × AOI relation matrix (where 1 = relation present). We then calculated the observed proportion of adjacent transitions that were relational. To estimate chance performance for that trial, we generated a trial-specific null distribution from 1,000 random sequences of the same length sampled from the set of unique objects visited on that trial, with immediate repeats disallowed. The relational score was defined as the z-scored difference between the observed relational-transition proportion and the mean of this null distribution. Thus, a relational score of 0 indicated chance-level relational scanning, while a positive relational score, for instance 1.5, indicated an observed relational scanning proportion 1.5 standard deviations above the trial's null baseline.

Free-recall responses were scored using an image-specific codebook pipeline. For each image, all non-empty free-recall responses were pooled and submitted to OpenAI's GPT-5.4 model (OpenAI, 2026), which extracted a deduplicated set of atomic memory concepts and classified them by content type, evidential status, and correctness (see Appendix C for the extraction prompt and category definitions). These image-level codebooks were then verified against the original stimulus image in a second GPT-5.4 vision pass, which corrected concept status against the visual evidence in the stimulus image, resolved contradictory concepts, and enforced the final concept taxonomy (see Appendix D for the vision verification prompt). The resulting verified codebook served as the source of truth for subsequent scoring.

The GPT-5.4 model was then used to score each participant's free-recall response against the verified codebook for that image, applying semantic matching rather than exact keyword matching (see Appendix E for the automated recall-scoring procedure). Only concepts marked as correct in the verified codebook were eligible to count as recalled. Recalled concepts were then aggregated by content type to derive three final memory outcomes: object recall (the sum of recalled object identities and attributes), relational recall (the sum of recalled action and spatial relations), and total recall (the sum of object and relational recall). Although scene-gist concepts were retained in the intermediate taxonomy, they were excluded from these final outcome measures. Low-scoring response–image pairs were additionally screened for possible wrong-image recalls, and responses identified as wrong-image recalls were excluded before analysis. To validate this automated scoring pipeline, the experimenter manually scored a 1/6 subset of participant responses. Agreement between automated and human scoring was strong at the category-count level (Intraclass Correlation Coefficient [ICC(2,1)] = .75 to .77; Pearson's $r$

= .88). In addition, agreement across two independent automated scoring runs was near-perfect at the individual-concept level, yielding 98.7% agreement (Cohen's $\kappa = .96$).

For the main analyses, raw recall counts were transformed into proportion scores separately for total, relational, and object recall by dividing each participant's score by the empirical maximum observed across all participants for that scene image and memory outcome. For example, if the highest object-recall count achieved by any participant for a given image was 10, then another participant who recalled 5 object details for that image would receive an object-recall proportion of .50. This placed all memory outcomes on a common 0 to 1 scale while controlling for between-image differences in recallability.

Relational scanning at encoding and retrieval was first tested against the chance-level baseline using one-sample t-tests. We then fitted linear mixed-effects models (LMMs) predicting total, relational, and object recall proportion from relational score, with separate models for encoding and retrieval. In these models, participant-level relational score was the main fixed effect of interest, while image-level relational score, fixation count, and salience at relational fixations were included as additional fixed effects. To separate individual viewing strategies from stimulus-driven effects, relational scores were group-mean centered per image. The resulting image-level relational scores were included as covariates, alongside fixation count and salience at relational fixations, with all continuous predictors standardized (z-scored) prior to model fitting. To test whether the effect of encoding relational score differed between relational and object recall, we fitted an additional exploratory LMM including memory type, encoding relational score, and the encoding relational score × memory type interaction as fixed effects. All LMMs were estimated using restricted maximum likelihood (REML) with a Nelder-Mead optimizer and included crossed random effects for participant and image, with random intercepts for each participant and each image to account for repeated observations and between-image differences in overall recallability.

## Results

Descriptive statistics are reported in Table 1.

**Table 1**

*Descriptive Statistics for Relational Score at Encoding and Retrieval and Recall Percentages*

| | M (SD) | 95% CI |
|---|---|---|
| Relational score (encoding) | 1.33 (0.31) | [1.20, 1.46] |
| Relational score (retrieval) | 0.19 (0.27) | [0.08, 0.30] |
| Total recall (%) | 55.34 (10.54) | [50.99, 59.69] |

| Relational recall (%) | 46.07 (10.11) | [41.90, 50.25] |
|---|---|---|
| Object recall (%) | 56.11 (10.60) | [51.74, 60.48] |

*Note*. CI = confidence interval. N = 25.

We first tested whether relational scanning exceeded the predefined chance baseline at encoding and retrieval. Normality of participant means was confirmed for both encoding and retrieval relational scores (Shapiro–Wilk *ps* ≥ .604). Relational score at encoding was significantly above baseline, $t(24) = 21.49$, $p < .001$, 95% CI [1.20, 1.46], $d = 4.30$, indicating strong relational scanning during initial viewing. Relational score at retrieval was also significantly above baseline, $t(24) = 3.56$, $p = .002$, 95% CI [0.08, 0.30], $d = 0.71$, indicating moderately strong relational scanning during blank-screen retrieval (see Figure 3).

**Figure 3**

*Relational Scores at Encoding and Retrieval*

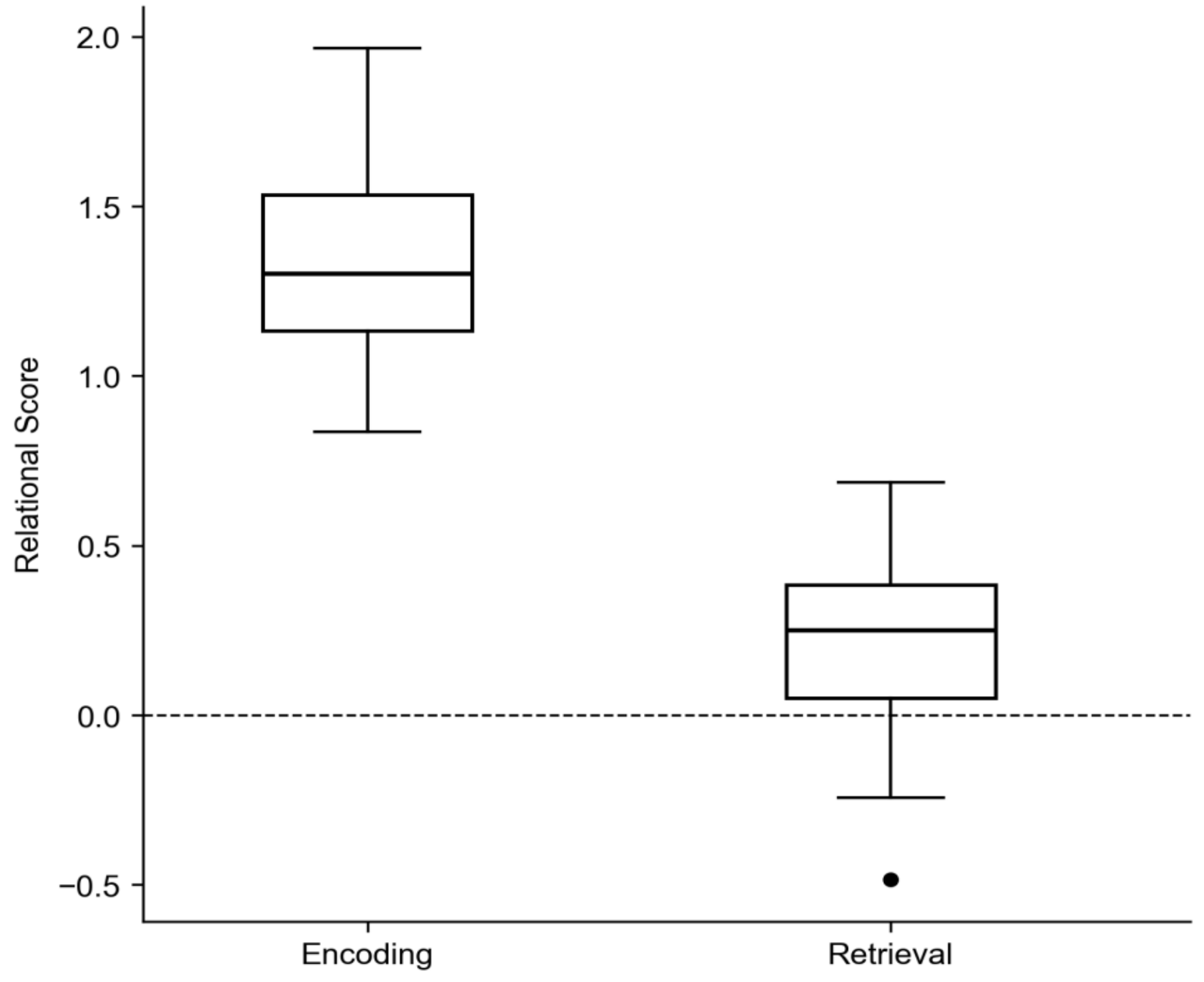

*Note.* The dashed line marks a relational score of 0, indicating chance-level relational scanning. In each boxplot, the box spans the interquartile range, the centre line marks the median, whiskers extend to values within 1.5 times the interquartile range, and individual points indicate outliers. N = 25.

To test whether relational scanning predicted subsequent memory, we fitted linear mixed-effects models predicting total, relational, and object recall from relational score, separately for encoding and retrieval. All models converged successfully, and preliminary diagnostic checks indicated no violations of normality, linearity, multicollinearity, or homoscedasticity. At encoding, greater relational score predicted higher total recall, $b$ = .024, 95% CI [.007, .040], $p$ = .006 (see Figure 4). The same positive pattern was also observed for relational recall, $b$ = .023, 95% CI [.003, .043], $p$ = .023, and object recall, $b$ = .021, 95% CI [.004, .039], $p$ = .018. Thus, when a participant scanned an image more relationally than was typical for that image at encoding, they subsequently recalled more content about it, including both relations and objects. Full model parameters including covariates are reported in Appendix A, Table 1, Panel A.

**Figure 4**

*Predicted Total Recall Percentage by Relational Score at Encoding*

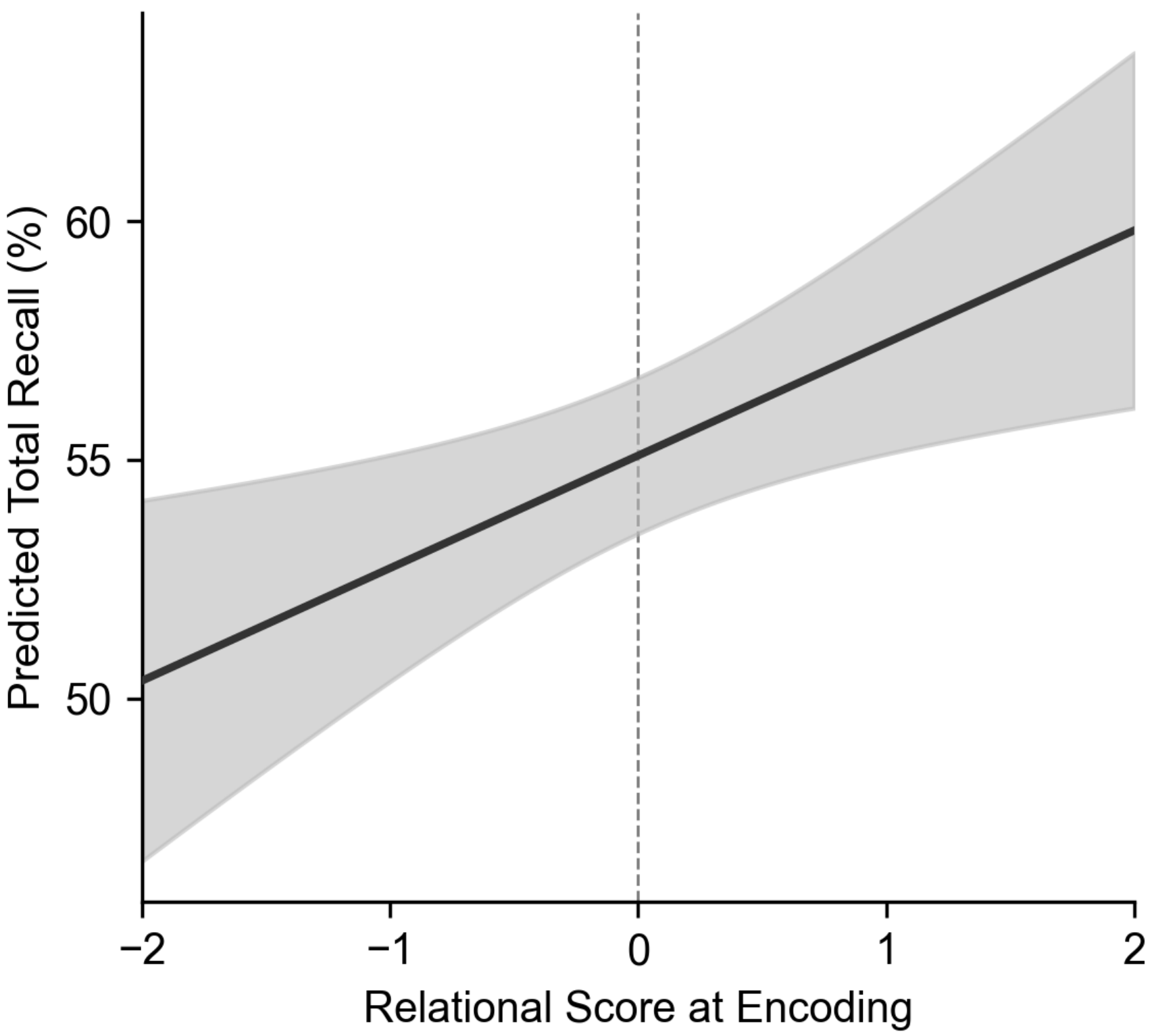

*Note*. The line shows the model-predicted association between relational score at encoding and total recall percentage. The dashed vertical line marks a relational score of 0, indicating chance-level relational scanning. Shaded regions represent 95% confidence intervals.

We next examined whether relational scanning at retrieval similarly predicted later memory. In contrast to the encoding results, greater relational score at retrieval did not predict total recall, $b$ = −.006, 95% CI [−.030, .018], $p$ = .629, relational recall, $b$ = −.002, 95% CI [−.031, .027], $p$ = .887, or object recall, $b$ = −.006, 95% CI [−.032, .019], $p$ = .612. Thus, although relational scanning at retrieval was above chance overall, it was not associated with better memory performance. Full model parameters including covariates are reported in Appendix A, Table 1, Panel B.

To determine whether the encoding effect differed between relational and object recall, we fitted an exploratory model including memory type and its interaction with encoding relational score. Relational recall was lower than object recall overall, $b = -.097$, 95% CI [−.115, −.079], $p < .001$. However, the interaction between encoding relational score and memory type was not significant, $b = .002$, 95% CI [−.016, .020], $p = .834$, showing no evidence that relational scanning at encoding benefited one type of recall more than the other (see Figure 5). Taken together, these findings suggest that greater relational scanning at encoding supported both relational and object recall, although object information was recalled more successfully overall. Full exploratory model parameters including covariates are reported in Appendix A, Table 2.

**Figure 5**

*Predicted Relational and Object Recall Percentage by Relational Score at Encoding*

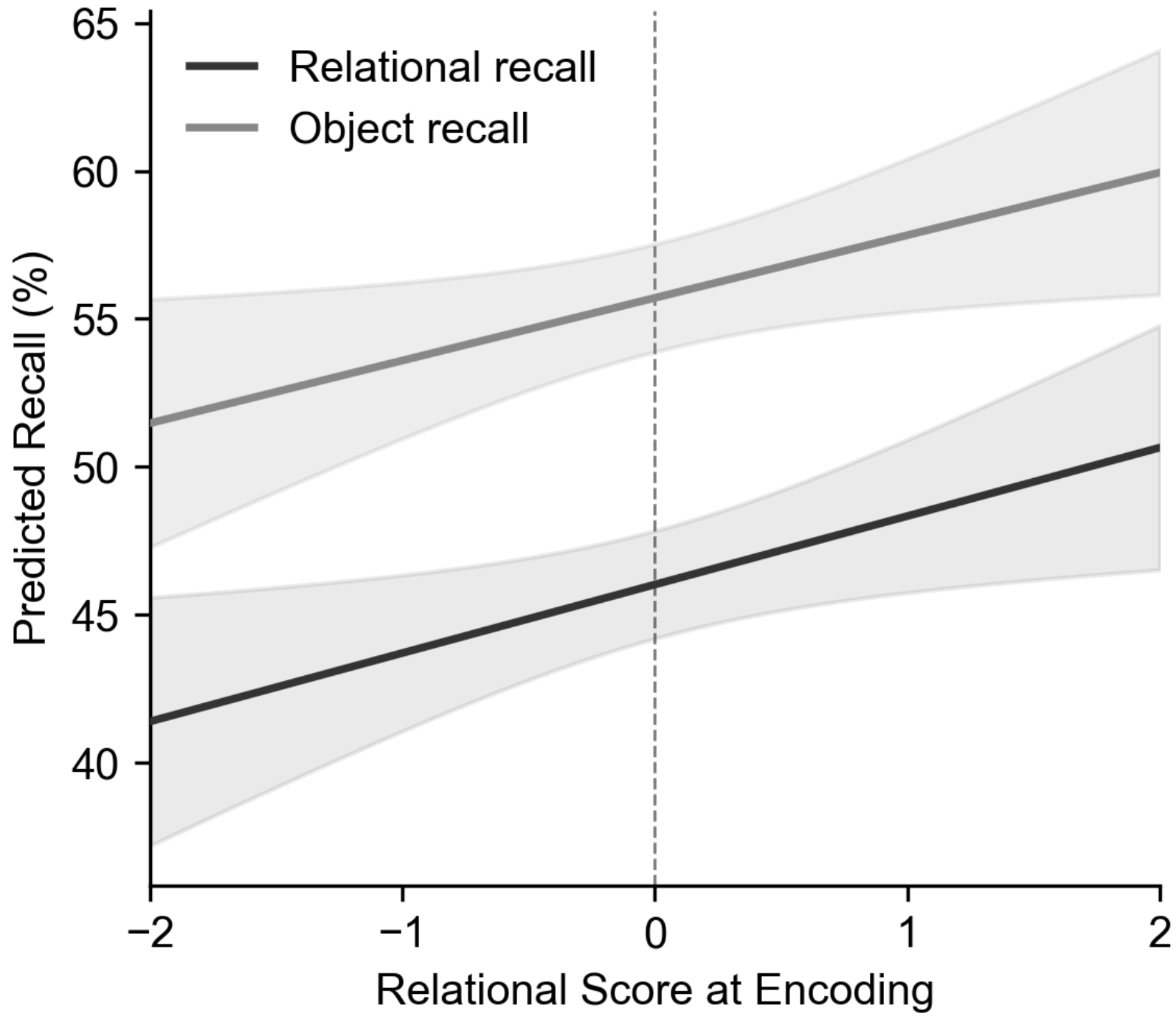

*Note.* Lines show model-predicted relational and object recall percentages as a function of relational score at encoding. The dashed vertical line marks a relational score of 0, indicating chance-level relational scanning. Shaded regions represent 95% confidence intervals.

## Discussion

The present study investigated whether relational gaze transitions support episodic memory for naturalistic scenes. Participants made relation-guided transitions significantly more often than expected by chance during both encoding ($M$ = 1.33, $SD$ = 0.31) and blank-screen retrieval ($M$ = 0.19, $SD$ = 0.27). This supports the first hypothesis, suggesting that relational structure is reflected in active visual exploration and can also be partially reinstated when the scene is no longer visible. However, this relational bias was only linked to memory success during encoding. Greater relational scanning during encoding predicted better subsequent recall for both object and relational details, whereas relational scanning during retrieval did not predict recall performance. Thus, the second hypothesis was supported for encoding but not retrieval. Together, these findings suggest that relational structure is expressed across both encoding and retrieval gaze but is most clearly linked to memory when it occurs during initial scene viewing.

The scanning effect observed at encoding suggests that relational structure systematically biases natural scene viewing. Previous attention-selection models have successfully explained which individual regions of a scene receive attentional priority based on visual salience and semantic meaning (Itti & Koch, 2000; Henderson & Hayes, 2017, 2018; Henderson et al., 2019). The present results extend these models by indicating that real-time visual sampling may also be structured by the relations among attended elements. In this way, gaze may provide a temporally ordered route through the relational layout of a scene. This adds an organisational level to accounts of visual processing, suggesting that eye movements reflect not only the selection of informative targets, but also the contextual linking of those targets within an unfolding event representation (Radvansky & Zacks, 2014; Zacks et al., 2007; Loschky et al., 2020). By identifying relational structure as a consistent influence on encoding gaze, these findings therefore provide a possible behavioural marker of how viewers navigate an event model during scene viewing.

This encoding pattern was also associated with later remembering. Greater relational scanning at encoding predicted better recall of both relational and object information, with comparable effects across both memory types. This pattern indicates that relational transitions supported memory for the scene, allowing individual objects to be embedded within a broader event structure. While relational memory theory posits that episodic remembering relies on binding elements together through their relations, it has historically lacked a clear account of the real-time viewing behaviours that facilitate this integration during naturalistic exploration (Konkel & Cohen, 2009; Eichenbaum, 2017). The present results bridge this gap. Mechanistically, transitions between related objects may allow recently sampled information to remain briefly active in visual working memory while a related partner is fixated, creating the temporal overlap necessary for binding (Hollingworth & Henderson, 2002; Hollingworth, 2004; Henderson & Hollingworth, 2003). Yet, because working memory and the binding process were not directly measured here, this mechanism remains a plausible but speculative interpretation. Nonetheless, the predictive power of encoding gaze points toward relational transitions as a measurable candidate for real-time contextual integration.

Despite this predictive relationship, the correlational design of the present study warrants a careful causal evaluation. While relational transitions may help construct a contextual scaffold for episodic recall, an alternative explanation is that efficient relational scanning indexes successful top-down scene comprehension. Under a schema-based view, an observer's active event model and prior knowledge could simultaneously guide relational eye movements and support subsequent encoding, without gaze transitions being the main cause of memory formation (Võ & Wolfe, 2013; Radvansky & Zacks, 2014). This leaves open whether relational gaze contributes to the construction of a novel episodic representation or instead reflects the rapid deployment of a pre-existing semantic schema. Addressing this distinction will require designs that separate relational gaze from prior scene knowledge, for example by comparing observers with different levels of domain knowledge or by testing whether specific relational transitions predict memory for the objects and relations they connect. Such work would clarify whether relational gaze contributes directly to episodic construction or is better understood as an index of scene comprehension.

Our findings during the retrieval phase clarify the relationship between overt eye movements and memory. During blank-screen retrieval, participants showed above-chance relational scanning, yet this behaviour did not predict free recall performance. This pattern contrasts with previous studies linking relation-guided gaze to successful retrieval (Chua et al., 2012; Hannula et al., 2007). A likely source of this discrepancy is the difference between target-constrained retrieval, such as cue-driven associative recognition and visual search tasks, and open-ended scene recall. When tasks require participants to identify a missing associate, verify a relation, or locate a target, reinstated spatial or relational information can directly constrain the memory search and guide attention toward a particular answer. By contrast, our present free recall task required participants to mentally reinstate an entire scene in preparation for broad, open-ended retrieval. In this context, shifting attention between remembered concepts may not require the eyes to physically revisit previously encoded spatial coordinates. Instead, retrieval may have depended more on scene gist and semantic organization, which can support a coherent verbal report without requiring gaze to replay the scene's relational layout (Polyn et al., 2009; Scholz et al., 2018). Taken together, these findings suggest the retrieval gaze effect is task dependent, serving a functional purpose primarily when memory search is constrained toward a specific target, while being less useful when participants reconstruct and report a whole scene from memory.

Methodologically, our SVG-based approach addressed the generalisability question raised by prior relation-specific studies. Previous work has shown that gaze can follow isolated types of relations, such as spatial links, functional actions, or social interactions (Hwang et al., 2011; Wu et al., 2014; Castelhano & Witherspoon, 2016; Federico & Brandimonte, 2019; Skripkauskaite et al., 2023). However, these studies leave open whether relation-sensitive scanning persists when many objects and relation types coexist within the same naturalistic scene. By combining object-level masks with comprehensive relation annotations, the current method allowed this broader test. The finding of above-chance relational scanning suggests that gaze can follow relational structure even amid the visual competition of complex scenes, extending prior evidence beyond isolated relation displays. However, the broad measure needed to test this general pattern also introduced an important interpretive limitation. Because our global relational score collapsed across annotated relations, it treated all relational transitions as equivalent. This is unlikely to capture the mnemonic structure of real scenes, where some relations may be more central to understanding the event than others (Radvansky & Zacks, 2014; Loschky et al., 2020). Social interactions provide one example of this problem. Interacting dyads are often remembered more effectively than non-interacting pairs (Vestner et al., 2022; Sun et al., 2025), suggesting that transitions between socially linked agents may carry greater mnemonic value than transitions between more peripheral objects. This illustrates the broader limitation that a relation's mnemonic value may depend not only on whether two objects are connected, but on how central that connection is to the scene's event structure. Thus, while our SVG approach suggests that relational scanning generalises to naturalistic scenes, it cannot determine which specific relational categories are most important for episodic encoding.

Future work should therefore move beyond a global relational score to a more granular, relation-specific analysis. One useful extension would be to categorise SVG relations into distinct domains, such as spatial, semantic, functional, and social links, to test whether certain types of relational scanning are prioritised during viewing and whether they predict recall more reliably. A complementary approach would be to weight relations by their conceptual centrality, using independent raters to judge how important each relation is to the scene's core event model (Radvansky & Zacks, 2014; Loschky et al., 2020). This contrast could help distinguish transitions between peripheral objects from those connecting the core agents, actions, and objects of the scene. Together, these extensions would help refine the theoretical case for relation-guided gaze by identifying which structural components may carry the greatest mnemonic value during episodic encoding.

In conclusion, this study identifies relation-guided gaze transitions as a behavioural marker of scene memory organization. Our results show that visual sampling follows relational scene structure during both active encoding and blank-screen retrieval. However, these transitions predict subsequent free recall only during initial encoding. This suggests that relational gaze may help integrate visual details into a cohesive event model mainly during memory formation, offering a framework to connect attentional selection with relational memory theory (Konkel & Cohen, 2009; Eichenbaum, 2017; Radvansky & Zacks, 2014). While our correlational design leaves open whether these transitions cause episodic binding or simply reflect scene comprehension, it helps narrow the gaze-memory link from general visual selection to the sequential organization of scene elements. By establishing a practical measure of relational scanning in complex scenes, this work provides a foundation for understanding how visual exploration contributes to episodic memory in real time.

## Data availability statement

All stimuli, analysis scripts, anonymized participant data, and outputs used in this study are available at: https://github.com/hugorydel/Relational_Scanpath_Analysis.

## Author contributions

H.R. conceived the study, designed and programmed the experiment, collected the data, performed the analyses, prepared the figures and tables, and wrote the first draft of the manuscript. A.K. supervised the project and contributed to the study design and revision of the manuscript. Both authors approved the final manuscript.

## Competing interests.

H.R. and A.K. declare no competing interests.

## Appendix

### Appendix A: Mixed-Effects Model Parameters

**Table 1**

*Fixed-Effects Parameter Estimates for H2 Mixed-Effects Models (Encoding and Retrieval)*

| | Total Recall | | Relational Recall | | Object Recall | |
|---|---|---|---|---|---|---|
| Predictor | b [95% CI] | p | b [95% CI] | p | b [95% CI] | p |
| **Panel A: Encoding phase (N = 695 observations)** | | | | | | |
| Relational score (encoding) | .024 [.007, .040] | .006 | .023 [.003, .043] | .023 | .021 [.004, .039] | .018 |
| Fixation count | .032 [.015, .049] | < .001 | .030 [.010, .049] | .004 | .026 [.009, .044] | .003 |
| Salience at relational fixations | -.025 [-.042, -.009] | .003 | -.020 [-.040, -.001] | .039 | -.018 [-.036, -.001] | .034 |
| Image-level relational score** | -.006 [-.022, .011] | .486 | .032 [.013, .052] | .001 | -.017 [-.034, -.000] | .048 |
| **Panel B: Retrieval phase (N = 334 observations*)** | | | | | | |
| Relational score (retrieval) | -.006 [-.030, .018] | .629 | -.002 [-.031, .027] | .887 | -.006 [-.032, .019] | .612 |
| Fixation count | .005 [-.020, .029] | .710 | .016 [-.014, .045] | .294 | .001 [-.024, .026] | .942 |
| Salience at relational fixations | -.000 [-.023, .022] | .987 | .018 [-.009, .045] | .185 | .006 [-.017, .029] | .619 |
| Image-level relational score** | -.024 [-.050, .001] | .064 | -.030 [-.061, .001] | .061 | -.019 [-.045, .008] | .168 |

*Note.* All continuous predictors were z-scored prior to model entry. Models included crossed random effects for participant and image. Estimates were made with a REML modelling approach, and a Nelder-Mead optimiser. *The low number of observations at retrieval results primarily from a large portion of trials which recorded fewer than 2 valid object transitions; these were removed from analysis. ** Image-level relational score controls for the baseline tendency of each image to elicit relational scanning across all participants.

**Table 2**

*Fixed-Effects Parameter Estimates for the Exploratory Dissociation Model*

| Predictor | b [95% CI] | p |
|---|---|---|
| Relational score (encoding) | .021 [.002, .040] | .027 |
| Memory type (relations vs. objects) | -.097 [-.115, -.079] | < .001 |

| Relational score (encoding) × Memory type | .002 [-.016, .020] | .834 |
|---|---|---|
| Fixation count | .028 [.012, .044] | .001 |
| Salience at relational fixations | -.019 [-.035, -.004] | .017 |
| Image-level relational score* | .008 [-.008, .024] | .349 |

*Note*. N = 1,390 observations, 25 participants. All continuous predictors were z-scored prior to model entry. The model included crossed random effects for participant and image. Estimates were made with a REML modelling approach, and a Nelder-Mead optimiser. * Image-level relational score controls for the baseline tendency of each image to elicit relational scanning across all participants.

## Appendix B: Automated Image Pre-Scoring Prompt

You are scoring ONE image using the rubric below. Output ONLY valid JSON matching the schema. No extra keys, no notes.

Step 0 — Core Interaction Set (max 3, MUST be UNIQUE + PRIMARY)
A "core interaction" is a visually obvious, intentional action between TWO DISTINCT visible entities, expressible as: Agent VERB Patient

PRIMARY-ACTION RULE (IMPORTANT — prevents secondary-action inflation)
- Prefer the highest-level, goal-defining event that best explains what's happening (the "main verb").
- Do NOT count a secondary/mechanism action if it is subordinate to a stronger action you already count in the same scene (e.g., holding/supporting/gripping/aiming/reaching/looking/standing-by) and it involves the same agent–patient pair.
- Examples (subordinate → exclude):
  - eating/biting/drinking → exclude "hold food/cup"
  - writing/signing/drawing → exclude "hold pen/paper"
  - tool-use (cutting/stirring/hitting/painting) → exclude "hold tool" and exclude "tool touches target" as a separate interaction
  - throwing/handing → exclude "hold object" if the transfer/throw is clearly depicted

Measure 1 — CIC (Core Interaction Count) [0–3]
- CIC = number of UNIQUE PRIMARY core interactions present in the scene.
- If there are 3+ unique primary interactions, set CIC = 3 and list the 3 most visually obvious.

Measure 2 — SEP (Separability) [0–2]
Score separability ONLY for the entities in your listed core_interactions.
- 0 = at least one required entity-pair is not reliably separable as AOIs (heavy occlusion/merging/tiny/one blob)
- 1 = separable but crowded/borderline

- 2 = clearly separable (distinct regions; fixations assignable with low ambiguity)

Measure 3 — DYN (Scene Dynamics) [0–2]
How “dynamic” is what’s happening between entities (vs mostly static/passive contact)?
- 0 = mostly static/passive (posing, sitting, holding with no clear action, looking, resting)
- 1 = some action but mild/slow/ambiguous
- 2 = clearly dynamic interaction(s) (force/motion/transfer/change is obvious)

Measure 4 — QLT (Image Quality) [0–1]
How reliable is this image for consistent action labeling and AOI marking?
- 0 = low quality / high noise
- 1 = clear / low noise

## Appendix C: Automated Gradebook Generation Prompt

You are an expert data annotator for a cognitive psychology study on visual memory.

Given a set of participant text descriptions of a single image, extract an exhaustive, deduplicated list of every distinct concept mentioned across all responses.

Rules for extraction:
- STRICT ATOMICITY: Do NOT combine distinct items into one entry.
  - NEVER output a number and a noun together. "Four girls" MUST be split into "four" (object_attribute) and "girls" a).
  - NEVER output an adjective and a noun together. "Green shirt" MUST be split into "green" (object_attribute) and "shirt" (object_identity).
  - The `object_identity` field must contain a noun ONLY. Adjectives and materials are forbidden in this field.
- DO merge surface variations of the same concept (e.g. "smelling", "sniffing", "smell", “sniff” → one entry).
- Extract only concepts that are genuinely memory-relevant; ignore filler phrases like "there is" or "I remember".

Classify each concept using this strict schema:

context (string):
  A brief description of what the concept modifies, relates to, or acts upon in the scene, so we know exactly how it was used. (e.g. "modifies the bag being held", "action performed by the woman", "reaching for the apple").

content_type (choose exactly one):
  object_identity  - Distinct physical entities / nouns (e.g. "woman", "dog", "surfboard")
  object_attribute - Descriptive properties: count, colour, size, material (e.g. "four", "green", "see-through")
  action_relation  - Dynamic interactions or motions (e.g. "running", "holding", "smelling")
  spatial_relation - Static physical positioning (e.g. "on the left", "inside", "behind")
  scene_gist       - Overarching setting or environment (e.g. "beach", "kitchen", "outdoors")

evidence_type (choose exactly one):
  literal      - Relies strictly on visible pixels; no synthesis required. Basic biology, geometry, colour, physical movement. (e.g. "woman", "red", "two", "touching")
  latent       - Pragmatic shortcut describing a highly probable visual reality. Synthesises visible cues into standard social or functional labels. Reasonable inference, but not directly pixel-readable. (e.g. "mother and daughter" inferred from age gap and proximity; "chef" from white coat; "smelling wine" from nose in glass)
  speculative  - Introduces narrative detail, internal mental states, or specific background knowledge not strongly afforded by the image. (e.g. "they are tourists",

"she is sad", "waiting for a bus")

status (choose exactly one):
  correct   - Concept is plausibly shared memory content for this image
  incorrect - Concept appears to be a hallucination or confabulation

source_phrases (list of short strings):
  For each participant response that mentions this concept, extract the 1-2 words immediately surrounding the concept that contextualize the match. Do NOT copy the full sentence.
  Examples: concept "dog" → ["big dog", "the dog", "a dog"] NOT ["I saw a big brown dog running"]
  concept "red" → ["red shirt", "red bag"] NOT ["she was wearing a red shirt"]
  Include one entry per distinct participant who mentioned this concept.

## Appendix D: Automated Codebook Editing Prompt

You are an expert visual data validator for a cognitive psychology study on visual memory.

You are given an image and a draft JSON Codebook of memory concepts extracted from participant text descriptions. Your job is to verify and edit the draft Codebook against the actual image, acting as the definitive ground truth.

CRITICAL EDITING RULES:

1. VERIFY STATUS
   Look at the image carefully. For each node:
   - If the concept is physically present or visually deducible (latent), set status to "correct".
   - If the concept is a participant hallucination — not in the image and not reasonably inferred — set status to "incorrect".
   - Override the draft status whenever the image contradicts it. Example: if participants recalled a "glass barrier" and you can see a reflective glass facade, mark it "correct" even if the draft said "incorrect".

2. SPLIT CONTRADICTIONS (The "Person" Fix)
   Review source_phrases for broad nodes where participants described mutually exclusive things (e.g., one said "blonde man", another said "woman" for the same person node).
   - Split these into distinct nodes.
   - Create one node for the visually verified truth (e.g., concept: "woman", status: "correct").
   - Create separate nodes for the hallucinated alternatives (e.g., concept: "man", status: "incorrect").
   - Each split node gets an appended suffix on the original node_id (e.g., "2387122_002a", "2387122_002b").

3. PRESERVE IDs
   - If you keep a node unchanged, preserve its original node_id exactly.
   - If you split a node into two or more, use suffix notation: original_id + "a", "b", "c" (e.g., "2387122_002a").
   - New nodes you add that have no original counterpart should use the pattern: "{StimID}_{next_available_index}".

5. PRESERVE FIELDS
   - Every output node must have exactly these fields: node_id, concept, context, content_type, evidence_type, status, source_phrases.
   - Do not add or remove fields.
   - source_phrases: for each participant who mentioned this concept, extract the 1-2 words immediately surrounding the concept that confirm the match. Do NOT copy full sentences.
     Examples: concept "dog" → ["big dog", "the dog"] NOT ["I saw a big brown dog running"]
     concept "red" → ["red shirt", "red bag"] NOT ["she was wearing a red shirt"]

- source_phrases for split or corrected nodes should contain only the phrases that map to that specific concept.

Return ONLY a valid JSON array of nodes. No preamble, no explanation, no markdown fences.

### Appendix E: Codebook Recall Scoring

You are an expert memory scorer for a cognitive psychology study on visual memory.

You are given a participant's free-text recall description of an image, and a Codebook of verified memory concepts for that image. Your task is to score whether the participant mentioned each concept in the Codebook.

SCORING RULES:

1. RECALL DECISION
   For each node in the Codebook, decide whether the participant's response contains this concept.
   Only return nodes where the concept WAS recalled. Omit nodes that were not recalled.
   Nodes absent from your response will automatically be scored as recalled=false.

2. SEMANTIC MATCHING (not keyword matching)
   Match on meaning, not exact words. Examples:
   - "kitty" matches concept "cat"
   - "typing" in the context of "typing on keyboard" matches concept "typing"
   - "sitting on something blue" matches concept "blue" AND concept "on" (spatial)
   - "furry animal" does NOT match concept "cat" — too vague, could be any animal
   Use judgement: the match must be specific enough that a human coder would agree.

3. MATCHED PHRASE
   For each recalled node, copy the shortest phrase from the response that triggered the match.

4. ONLY SCORE CORRECT NODES
   Nodes with status "incorrect" in the Codebook are hallucinated concepts — do not return them
   even if the participant mentions them.

5. PRESERVE NODE IDs
   Use node_id values exactly as they appear in the Codebook. Do not add or invent node IDs.

Return ONLY a valid JSON array of recalled nodes. No preamble, no explanation, no markdown fences.
Each element must have exactly: node_id (string), matched_phrase (string).
If no nodes were recalled, return an empty array: []

### Appendix F: Excerpted Outputs from Codebook

```
[
 {
   "node_id": "2374670_001",
   "concept": "vehicle",
```

"context": "main transport object in the scene being driven and spraying liquid",
"content_type": "object_identity",
"evidence_type": "literal",
"status": "correct",
"source_phrases": [
"spray vehicle",
"red truck",
"red van",
"spraying truck",
"vehicle spraying"
]
},
{
"node_id": "2374670_002",
"concept": "red",
"context": "modifies the vehicle",
"content_type": "object_attribute",
"evidence_type": "literal",
"status": "correct",
"source_phrases": [
"red truck",
"red van",
"red spray",
"was red"
]
},
{
"node_id": "2374670_003",
"concept": "person",
"context": "person on or in the vehicle",
"content_type": "object_identity",
"evidence_type": "literal",
"status": "correct",
"source_phrases": [
"a man",
"guy on",
"man standing",
"guy in"
]
},
[...55 more concept nodes hidden…]]

**Appendix G**

**Table 1**

*Representative Encoding Engagement Questions Used Across First and Second Scene Viewings*

| Cue word | Viewing | Question | Response options | Correct answer |
|---|---|---|---|---|
| wine | 1 | What are the men doing with the wine? | clinking; pouring; smelling; sipping | smelling |
| wine | 2 | What are the men sitting on in the scene? | couch; chairs; bench; table | couch |
| blender | 1 | Which part of the blender is the woman holding? | glass; handle; lid; base | lid |
| blender | 2 | What is the man doing to power the blender? | pedaling a bicycle; turning a crank; pressing a button; pulling a lever | pedaling a bicycle |
| polar bear | 1 | Where is the polar bear located in the scene? | on rocks; on ice; in water; on grass | in water |
| polar bear | 2 | What is the polar bear doing next to the enclosure edge? | pressing on glass enclosure; swimming away from glass enclosure; diving underwater; sitting on snowy ground | pressing on glass enclosure |
| suitcase | 1 | What are the men doing with the suitcases? | stacking them; carrying them; pulling them; opening them | carrying them |
| suitcase | 2 | Where are the men? | walking along a train platform; walking into a hotel lobby; walking under a covered walkway; walking in an airport terminal | walking under a covered walkway |